\documentclass[12pt]{article}
\usepackage{amsmath}
\usepackage{amssymb}
\usepackage{amsfonts}
\usepackage{multirow}
\usepackage{mathrsfs}
\usepackage{bbm}
\usepackage{graphicx}
\usepackage{float}
\usepackage{subfigure}
\usepackage{diagbox}
\usepackage{rotating}
\usepackage{multirow,color}
\usepackage{amsbsy}
\usepackage{latexsym}
\usepackage{comment}
\usepackage{natbib}
\renewcommand{\baselinestretch} {1.5}
\makeatletter 
\def\singlespace{\def\baselinestretch{1}\@normalsize}

\@addtoreset{equation}{section}
\renewcommand{\theequation} {\arabic{section}.\arabic{equation}}

\newtheorem{theorem}{Theorem}
\newtheorem{lemma}{Lemma}
\newtheorem{corollary}{Corollary}

\newtheorem{proposition}{Proposition}

\def\hat{\widehat}
\def\tilde{\widetilde}
\def\bar{\overline}

\def\bs{\boldsymbol}

\begin{document}
\title{An Empirical Bayes Method for Chi-Squared Data}
\author{Lilun Du and Inchi Hu\footnote{\textit{Address for correspondence}: Inchi Hu, Department of Information Systems, Business Statistics and Operations Management, Hong Kong University of Science and Technology, Hong Kong. E-mail: imichu@ust.hk} \\
{\it Hong Kong University of Science and Technology, Hong Kong} }
\date{\today}
\maketitle{}

\begin{abstract}
\noindent
In a thought-provoking paper, \cite{Efron:2011} investigated the merit and limitation of an empirical Bayes method to correct selection bias based on Tweedie's formula first reported in \cite{Robbins:1956}. The exceptional virtue of Tweedie's formula for the normal distribution lies in its representation of selection bias as a simple function of the derivative of log marginal likelihood. Since the marginal likelihood and its derivative can be estimated from the data directly without invoking prior information, bias correction can be carried out conveniently. We propose a Bayesian hierarchical model for chi-squared data such that the resulting Tweedie's formula has the same virtue as that of the normal distribution. Because the family of noncentral chi-squared distributions, the common alternative distributions for chi-squared tests, does not constitute an exponential family, our results cannot be obtained by extending existing results. Furthermore, the corresponding Tweedie's formula manifests new phenomena quite different from those of the normal distribution and suggests new ways of analyzing chi-squared data.
\end{abstract}

\newpage
\section{Introduction}\label{Sec-1}
In this paper, we take the chi-squared test to be any statistical test such that the test statistic under the null hypothesis is approximately chi-squared distributed. Pearson's chi-squared test represents an important subclass. Other examples include Fisher's exact test, the Cochran-Mantel-Haenszel test, McNemar's test for contingency tables, Turkey's test for additivity in the analysis of variance, the portmanteau test in time-series analysis, and Wald's test and likelihood ratio test in general statistical modelling etc. Indeed, the chi-squared test is one of the most widely used statistical hypothesis tests. Among other objectives, it can be applied to test goodness of fit, homogeneity, and independence etc.

Suppose we conduct a large number of chi-squared tests simultaneously. Based on these test results, not only would we like to know which test is significant after adjustment for multiplicity, but also the effect size of the significant test results. Large scale studies and the associated compound decision problems of this kind have motivated a revival of empirical Bayes methods; see \cite{Efron:2010}.

Our approach to the issue raised above hinges on Tweedie's formula based on a Bayesian hierarchical model for chi-squared data, which is then employed to construct posterior intervals of the effect size. The Bayesian hierarchical model has a long history in empirical Bayes literature. \cite{Robbins:1956} contained several remarkable Bayesian estimation formulae under such models, which were referred to as Tweedie's formula by \cite{Efron:2011}. The superiority of empirical Bayes estimates derived from such formulas and their dominance over maximum likelihood estimates are part of the profound study in \cite{Brown:1971} and \cite{Stein:1981}. 
More recent works include \cite{YL:2005}, \cite{Jiang:Zhang:2009}, \cite{Muralidharan:2010}, \cite{Brown_etal_2013}, \cite{FJS:2017}, and \cite{Weinstein_etal_2018}.

In this paper, we assume that the chi-squared statistic $\{X_i\}_{i=1}^{m}$ for a large number $m$ of chi-squared tests are generated from the following Bayesian hierarchical model
\begin{equation}\label{cpdmodel}
 J_i | \lambda_i \sim {\rm Poisson(\lambda_i/2)}, \;X_i | J_i\sim\chi^2_{k+2J_i},
\end{equation}
where the noncentrality parameters $\lambda_i$'s are {\it i.i.d.} draws from an unspecified prior distribution
\[\lambda_i\sim G(\lambda), \;i=1,\cdots, m.\]
Here $\lambda_i$ and $J_i$, $i=1,\cdots, m$ are not observable and $X_i$, $i=1,\cdots, m$ are observable. The goal is to estimate the noncentrality parameters $\lambda_i$, $i=1,\cdots, m$ based on the sample data $\{X_i\}_{i=1}^m$ as a compound estimation problem.

We will show later that the posterior mean $E(\lambda|X=x)$ has a close-form representation in the marginal density and its derivatives like Tweedie's formula in the empirical Bayes literature. It is notable that Tweedie's formula is previously known to hold only in the exponential family, while ours is outside the exponential family. With the new Tweedie's formula, we can estimate the Bayes rule $E(\lambda|x)$ from the sample data and make progress on the corresponding compound estimation problem, which typically involves the estimation of all $\lambda_i$'s. However, in large scale studies, it is quite common to select a subset of test statistics, such as $K$ largest values of $\{X_i\}_{i=1}^{m}$, for followup study. Such a selected subset is subjected to selection bias as perceptively explained in \cite{Efron:2011}.

The Bayes rule is free from selection bias which is nicely explained in \cite{Dawid:1994}, \cite{Senn:2008}, and \cite{LD:2016}. \cite{Efron:2010, Efron:2011} therefore advocated an empirical Bayes procedure for large scale inference, incorporating a plug-in estimate of the marginal likelihood and its derivatives in Tweedie's formula to correct for selection bias. The procedure does not rely on the particular form of selections and works for selection of any kind, including no selection. For the same reason, selection is not a prerequisite of our method. 


It is possible to transform the chi-squared values into z-values, then use Tweedie's formula for normal distributions, following \cite{Efron:2010, Efron:2011}, to overcome selection bias. The merit of this approach is to save the trouble of developing a new theory for chi-squared tests. On the other hand, in the process of transformation, we lose the intrinsic meaning of the noncentrality parameter of chi-squared distribution and have to interpret the chi-squared data in normal-distribution terms. In this paper, we develop a separate theory so that the chi-squared data can be interpreted in their own right.

Our contribution is twofold. First, we formulate a somewhat unexpected model showing that Tweedie's formula can hold outside the exponential family. Secondly, we introduce new statistical tools to carry out a type of post-selection inference for chi-squared data. The rest of the paper is organised as follows. In Section~\ref{Sec-2}, we present a Bayesian hierarchical model, which can accommodate data from a large number of chi-squared tests. In Section~\ref{Sec-3}, we show that it is possible to derive explicit formulae for the posterior mean and variance of the noncentrality parameter from an unspecified prior, which are then employed to construct posterior intervals for the effect size. Section~\ref{Sec-3a} explains how to estimate log-density derivatives in the posterior mean and variance. Section~\ref{Sec-4} suggests ideas to interpret the noncentrality parameter estimates. Section~\ref{Sec-5} contains a simulation study on a sparse model with interaction effects, exploring possible applications to variable selection for high-dimensional data. Two real data examples are presented in Section~\ref{Sec-6}. Some concluding remarks are given in Section~\ref{Sec-7}. The proofs are in the appendix.

\section{From Problem Description to Statistical Modeling}\label{Sec-2}
\subsection{A Motivating example}

Suppose subjects under study come from several groups. The group membership is decided by either innate or acquired attributes of the subject. For instance, they can be patients subjected to different treatments or individuals from different ethnic groups etc. We would like to test homogeneity among groups with respect to some response variables. In particular, consider the following simple model for the expression level of gene $i$
\begin{equation}\label{lm}
{\bf Y}_i = {\bf X}{\bs \beta}_i+{\bs \epsilon}_i,
\end{equation}
where ${\bf Y}_i$ is the vector of expression levels of gene $i$ for all subjects in the study, ${\bf X}$ the matrix describing to which group each subject belongs, ${\bs \beta}_i$ the vector of parameters determining gene $i$ mean expression level for each group, and ${\bs \epsilon}_i$ are random errors.

To find out which among a large number of genes generate inhomogeneous expression levels for subjects from different groups, we test the null hypothesis of homogeneity, which is a test of equality of several group means. This is a linear hypothesis testing problem; see e.g. Chapter 7 of \cite{Lehmann:1986}. The likelihood ratio test is also the uniform most powerful (UMP) invariant test. The test statistic has $F$-distribution under the null hypothesis and is well approximated by chi-squared distribution when the denominator degrees of freedom (essentially the number of subjects in the study) is large. This situation brings one face to face with a large number of chi-squared tests, one for each gene, simultaneously.

After conducting a large number of chi-squared tests, such as those mentioned above, one selects a subset of large chi-squared statistic values to estimate the effect sizes. The usual parameter estimates based on selected chi-squared values are subjected to selection bias and thus misleading. We next introduce a Bayesian hierarchical model to address the selection bias issue amid a large number of chi-squared tests.

For easy reference, we record here the chi-squared density function with $k$ degrees
\[f_k(x)=\frac{1}{2^{k/2}\Gamma(k/2)}x^{k/2-1}e^{-x/2},\]
where $\Gamma(a)=\int_0^\infty t^{a-1}e^{-t}dt$ is the celebrated gamma function.
It is known that the non-central chi-squared distribution with noncentrality parameter $\lambda$ can be written as the Poisson mixture of chi-squared densities as follows
\[f_{k,\lambda}(x)=\sum_{j=0}^{\infty}e^{-\lambda/2}\frac{(\lambda/2)^j}{j!}f_{k+2j}(x). \]
Let $G(\lambda)$ be the prior distribution function over $[0, \infty)$. Whenever necessary, we assume that the prior density exists and denote it by $g(\lambda)$. The marginal density function is
$g_k(x)=\int_0^{\infty}f_{k,\lambda}(x)dG(\lambda)$
and the posterior density equals
$g_k(\lambda|x)=f_{k,\lambda}(x)g(\lambda)/g_k(x)$.

We encapsulate the preceding results in a Bayesian hierarchical model as follows
\begin{equation}\label{model}
\lambda \sim G(\lambda),\; J | \lambda \sim{\rm Poi}(\lambda/2),\; X | J \sim \chi_{k+2J}.
\end{equation}
First draw the noncentrality parameter $\lambda$ from the unspecified prior distribution $G(\lambda)$. Next generate a non-negative integer $J$ according to a Poisson distribution with mean $\lambda/2$.  Then sample from the chi-squared population with $k+2J$ degrees of freedom. Thus condition on $\lambda$, $X$ has a noncentral chi-squared distribution with noncentrality parameter $\lambda$ and $k$ degrees of freedom. Repeating (\ref{model}) $m$ times, we arrive at (\ref{cpdmodel}), a generative model for data from a large number of chi-squared tests.

\subsection{Noncentrality parameter as effect size of chi-squared test}\label{Sec-2.1}
Chi-squared statistics often arise as the sum of square independent normal variates $Y_i\sim N(\mu_i, 1), ~i=1,\cdots, k$. For instance, dropping the index $i$ in the linear model (\ref{lm}) to test the null hypothesis $H_0:{\bf A}\bs\beta = \bs{0}$ for ${\bf A} \in \Re^{k\times p}$, the $F$-statistic up to a constant equals
\[\tilde{T} = \{\mathbf{A}\hat{\bs{\beta}}\}^\top
\{\mathbf{A}(\mathbf{X}^\top\mathbf{X})^{-1}\mathbf{A}^\top \}^{-1}
\{\mathbf{A}\hat{\bs{\beta}}\}/\hat{\sigma}^2,
\]
where $\hat{\bs{\beta}}=(\mathbf{X}^\top\mathbf{X})^{-1}\mathbf{X}^\top \mathbf{Y}$ and $\hat{\sigma}^2$ are the maximum likelihood estimates of $\bs{\beta}$ and $\sigma^2$ the variance of random errors, respectively. The asymptotic distribution of $\tilde{T}$ is the same as $T=\sum_{i=1}^k Y_i^2$ with $\sum_{i=1}^k \mu_i^2 ={(\bf A}\bs\beta)^\top\{{\bf A}({\bf X}^\top {\bf X})^{-1}{\bf A}^\top\}^{-1}{\bf A}\bs\beta/\sigma^2$.

Under the alternative hypothesis $H_1: {\bf A}\bs\beta \neq \bs{0}$, $T$ has noncentral chi-squared distribution with noncentrality parameter $\lambda=\sum_{i=1}^k\mu_i^2$. The distribution of $T$ depends on $\mu=(\mu_1,\cdots, \mu_k)^\top$ only through the noncentrality parameter $\lambda$ due to rotational invariance of the normal distribution. If we view $\mu_i$ as the effect size of $Y_i$, then the noncentrality parameter equals the squared distance between the null and non-null mean vectors.

The noncentral chi-squared distributions are {\it stochastically increasing} in the noncentralily parameter $\lambda$; see e.g. \cite{van der Vaart:1998}. Hence the power of a chi-squared test is an increasing function of $\lambda$, which implies that the test with higher underlying noncentrality parameter has higher power. Sometimes, this monotone relationship can be quite simple to describe using  $\lambda$. \cite{CR:1987} obtained the following approximation of a noncentral chi-squared probability by a central one for small $\lambda$
\[P\{\chi_k(\lambda)>x\}\approx P\{\chi_k>\frac{x}{1+\lambda/k}\}.\]

In brief, common alternative hypotheses for chi-squared tests and rotational invariance of normal distributions lead to noncentral chi-squared distributions with noncentrality parameter $\lambda$, which has intrinsic geometrical meaning. Moreover, the noncentrality parameter ranks chi-squared tests according to their powers. These facts together provide compelling reasons to adopt the noncentrality parameter $\lambda$ as the {\it effect size} of chi-squared tests. 

\section{Tweedie's Formula for Chi-squared Data}\label{Sec-3}
The main result of this section concerns the posterior mean of model (\ref{model}).
\begin{theorem}\label{pm_thm}
Under model (\ref{model}) with $k\geq 2$ and an unspecified prior distribution $G(\lambda)$, the posterior mean of the effect size $\lambda$ given the observed value $X=x>0$ can be calculated from the marginal log-density derivatives according to
\begin{equation}\label{pmean2}
E_k(\lambda|x)=\left[(x-k+4)+2x[\frac{2l_k^{''}(x) }{1+2l_k^{'}(x)}+l_k^{'}(x)]\right]
\left[1+2l_k^{'}(x)\right],
\end{equation}
\end{theorem}
where $l'_k(x)=g'_k(x)/g_k(x)$ and $l''_k(x)=\{\log g_k(x)\}''$ equal the first and second derivatives of the marginal log-density, respectively.

\subsection{Preliminary results for posterior mean}
To help appreciate the implications of Theorem \ref{pm_thm}, we will go through a few preliminary results. When a chi-squared statistic $X=x$ is observed under model (\ref{model}), then
\begin{eqnarray}\label{pmean0}
E_k(\lambda|x)
&=&\frac{\int_0^{\infty}\lambda\sum_{j=0}^{\infty}e^{-\lambda/2}\frac{(\lambda/2)^j}{j!}f_{k+2j}(x)dG(\lambda)}{g_k(x)} \nonumber \\
&=&\frac{\int_0^{\infty}\sum_{j=0}^{\infty}2(j+1)e^{-\lambda/2}\frac{(\lambda/2)^{j+1}}{(j+1)!}f_{k+2j}(x)dG(\lambda)}{g_k(x)}
\nonumber \\
&=&\frac{\int_0^{\infty}\sum_{j=0}^{\infty}2je^{-\lambda/2}\frac{(\lambda/2)^{j}}{j!}f_{k-2+2j}(x)dG(\lambda)}{g_k(x)}
\nonumber \\
&=& \frac{E_{k-2}(2J|x)}{g_{k}(x)/g_{k-2}(x)},
\end{eqnarray}
where $E_{k-2}(2J|x)$ is calculated under $(k-2)$ null degrees of freedom.
To proceed further, we need

\begin{lemma}\label{Lemma-1}
The following relationship holds between the marginal density and its derivative under model (\ref{model})
\[g'_k(x):=\frac{d}{dx}g_k(x)=\frac{1}{2}[g_{k-2}(x)-g_k(x)],\]
or equivalently $g_{k-2}(x)=2g'_k(x)+g_k(x)$.
\end{lemma}
{\it Proof}\;\; Taking the derivative of chi-squared density function
\[\frac{d}{dx}f_{k+2j}(x)=\frac{1}{2}[f_{k-2+2j}(x)-f_{k+2j}(x)]\]
and integrating with respect to $\lambda$ and $j$ to obtain $g_k(x)$ yield the desired result. \hfill $\Box$

Applying Lemma~\ref{Lemma-1} and (\ref{pmean0}), we arrive at the following result.
\begin{proposition}\label{Proposition_1}
Assume that data are generated from model (\ref{model}) with $k\geq 2$ and an unspecified prior distribution $G(\lambda)$, the posterior mean of noncentral parameter $\lambda$ given $x>0$
\begin{equation}\label{pmean1}
E_k(\lambda|x)=\frac{E_{k-2}(2J|x)}{g_k(x)/g_{k-2}(x)}=E_{k-2}(2J|x)[1+2\frac{g'_k(x)}{g_k(x)}]
=E_{k-2}(2J|x)(1+2l'_k(x)),
\end{equation}
where $l'_k$ is the derivative of marginal log likelihood with $k$ degrees of freedom.
\end{proposition}

We can view (\ref{pmean1}) as a preliminary Tweedie's formula for chi-squared distribution. It bears considerable resemblance to Tweedie's formula for the normal distribution
\begin{equation}\label{normal}
E(\mu|z)=z+\sigma^2l'(z).
\end{equation}
The most visible feature in Tweedie's formula for normal distributions (\ref{normal}) and chi-squared distributions (\ref{pmean1}) is that both depend on marginal log likelihood in an essential way. Since $E(2J|\lambda)=\lambda$, $2J$ reflects the effect size in (\ref{model}). At the risk of terminology abuse, call $E_{k-2}(2J|x)$ the `pseudo-observed' effect size, as (\ref{model}) has $2J$ one step closer to the observed value $x$ than $\lambda$. The variance of chi-squared random variable is twice the mean, which explains the `2' in $1+2l'_{k}(x)$. Altogether we see remarkable similarity between Tweedie's formula for the normal and the chi-squared distributions, matching everywhere except for the curious appearance of $k-2$ instead of $k$ as the null degrees of freedom in $E_{k-2}(2J|x)$.

\vspace{0.3in}
\begin{figure}[ht]
\centering
\vspace{-0.2cm}
\begin{tabular}{cc}
\includegraphics[width=0.9\textwidth]{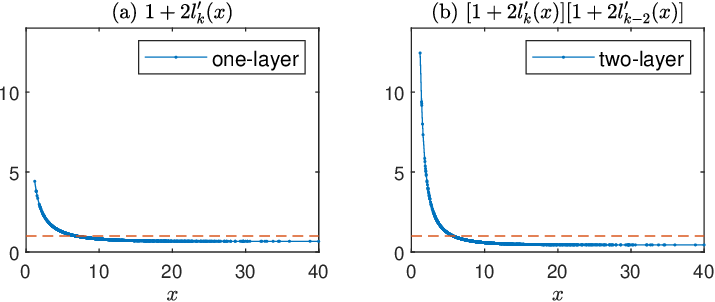}
\end{tabular}
\caption{\textsl{(a) Plot of one-layer multiplier $1+2l'_k(x)$; (b) Plot of two-layer multiplier $[1+2l'_{k-2}(x)][1+2l'_k(x)]$. In both plots, $k=7$ and $g(\lambda)=(1/4)e^{-\lambda/4}, \lambda>0$. }}\label{onelayer}
\vspace{-0.2cm}
\end{figure}

Figure~\ref{onelayer}(a) indicates that the multiplier $1+2l'_k(x)$ of the pseudo-observed value $E_{k-2}(2J|x) $ in (\ref{pmean1}) is bigger than one for small $x$, implying the posterior mean is larger than the observed value $x$. For larger $x$ value, it is smaller than one, implying the posterior mean is smaller than observed value $x$. For $x$ near $k$, the posterior mean is close to the observed value $x$. All these are anticipated bias-correction outcomes of the posterior mean.

Proposition~\ref{Proposition_1} expresses the posterior mean in terms of $E_{k-2}(2J|x)$, the expected effect-size degrees of freedom. We now study how to estimate it from the data. The resulting estimates are inaccurate and we shall not use them for bias correction. Nevertheless, these estimates provide insight into $E_{k-2}(2J|x)$ as an important part of the posterior mean.

According to (\ref{model}), $X=x$ in $E_{k-2}(2J|x)$ is selected from a chi-squared population with $k-2+2J$ degrees of freedom. If we employ maximum likelihood estimation, then $x={\rm mode}\{X\}= \max\{ k-4+2J, 0 \}$, which leads to
\begin{equation}\label{xhat1}
 \hat{2J}_{k-2}=(x-k+4)_+,
\end{equation}
where $a_+=\max\{0,a\}$. If one prefers the method of moments, then $x=E_{k-2}(X)=k-2+2J$ and the corresponding estimate is
\begin{equation}\label{xhat2}
\tilde{2J}_{k-2}=(x-k+2)_+.
\end{equation}
Both methods estimate the effect-size degrees of freedom by soft-thresholding. It appears that the presence of $E_{k-2}(2J|x)$ in (\ref{pmean1}) carries a shrinkage effect on the observed value $x$.

\subsection{Tweedie's formula for posterior mean}
While being a Tweedie-type of result, (\ref{pmean1}) is not suitable for selection-bias correction, since $E_{k-2}(2J|x)$ is not directly estimable by sample data. Although we can estimate $E_{k-2}(2J|x)$ by (\ref{xhat1}) and (\ref{xhat2}), both estimates are unsatisfactory and their accuracy does not improve even with infinite amount of data. Here we present a formula for $E_{k-2}(2J|x)$ so that the Tweedie's formula (\ref{pmean2}) of Theorem \ref{pm_thm} can be obtained
\begin{equation}\label{e2j}
E_{k-2}(2J|x)=2x[\frac{2l_k^{''}(x) }{1+2l_k^{'}(x)}+l_k^{'}(x)]+(x-k+4).
\end{equation}
The proofs of (\ref{e2j}) and Theorem \ref{pm_thm} are given in the appendix.

The second term, $x-k+4$, on the right hand side of (\ref{e2j}) matches (\ref{xhat1}), which to some extent explains why (\ref{xhat1}) is inaccurate. Comparing to (\ref{e2j}), (\ref{xhat1}) lacks a bias correction term like the first term on the right hand side of (\ref{e2j}), which `borrows strength' from nearby observed values, and thus exposes the primitive nature of (\ref{xhat1}) as an estimate of $\lambda$.

The formula (\ref{pmean2}) can be employed to estimate the posterior mean after estimating the 1st and 2nd derivatives of log-density by the sample data. The other useful expression of the posterior mean is given by the following corollary.
\begin{corollary}
Under the same conditions as Theorem \ref{pm_thm}, the posterior mean of the model (\ref{model}) can also be written as
\begin{eqnarray}\label{pmean2a}
E_k(\lambda|x)&=&[(x-k+4)+2xl'_{k-2}(x)][1+2l'_k(x)] \nonumber \\
&=&x[1+2l'_{k-2}(x)][1+2l'_k(x)] - (k-4)[1+2l'_k(x)].
\end{eqnarray}
\end{corollary}
Equation (\ref{pmean2a}) reveals that the posterior mean achieves bias correction in two steps, which is totally new to Tweedie's formula for normally distributed data. The first step is a two-layer multiplicative adjustment of $x$, followed by a deduction by a reduced null degrees of freedom times the first layer of multiplicative adjustment. The next lemma is useful in comparing the two layers of multiplicative adjustment.
\begin{lemma}\label{Lemma-2}
If the marginal density $g_k$ is log concave, then the derivative of marginal log likelihood $l'_k(x)>l'_{k-2}(x)$ for each $x>0$.
\end{lemma}
{\it Proof.}\;\; Since
\[2l'_k(x)+1=g_{k-2}(x)/g_k(x),\]
it is sufficient to prove that $g_{k-2}(x)/g_k(x)>g_{k-4}(x)/g_{k-2}(x)$ for each positive $x$ value.

Log concavity of $g_k(x)$ implies that $l''_k(x)<0$. Using Lemma~\ref{Lemma-1}, we can show that
\[l''_k(x)=\frac{1}{4}\left[\frac{g_{k-4}(x)}{g_k(x)}-\left(\frac{g_{k-2}(x)}{g_k(x)}\right)^2\right]<0,\]
which is equivalent to $g_{k-2}(x)/g_k(x)>g_{k-4}(x)/g_{k-2}(x)$. The proof is completed. \hfill $\Box$


Assuming log concavity, Lemma~\ref{Lemma-2} implies that the second layer of correction, $l'_{k-2}(x)$, produces smaller adjustment than the first layer $l'_k(x)$ for small $x$ value because $l'_k(x)>l'_{k-2}(x)>0$ for $x$ close to zero. For large $x$ value, however, the second layer produces larger adjustment than the first layer since $0>l'_k(x)>l'_{k-2}(x)$ for $x$ far away from zero. 


The two-layer adjustment is plotted in Figure~\ref{onelayer}(b) and basically the same as one layer: pulling up small value and pushing down large value of $x$. The adjustment at both ends are obviously magnified by incorporating the second layer.

\vspace{0.2in}
\begin{figure}[ht]
\centering\vspace{-0.2cm}
\includegraphics[width=0.6\textwidth]{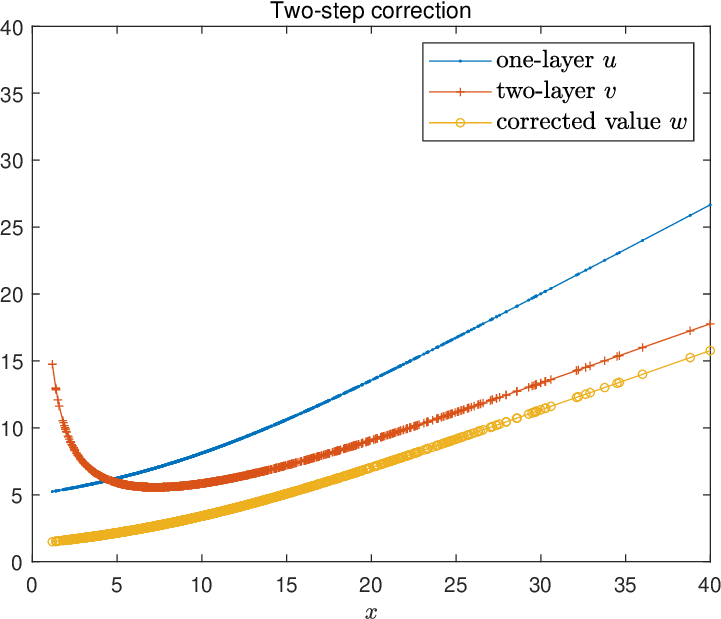}
\caption{\textsl{Bias correction effect of the posterior mean with $k=7$ and $g(\lambda)=(1/4)e^{-\lambda/4}$; one layer $u=x(1+2l'_k(x))$; two-layer $v=u(1+2l'_{k-2}(x))$; $E_k(\lambda|x)= w=v-(k-4)u/x$.} }\label{correctedvalue}
\vspace{-0.2cm}
\end{figure}

The bias correction effect of the posterior mean is shown in Figure \ref{correctedvalue}. With only two layers of multiplicative adjustment, the resulting values $v$ are too large for small $x$ values and have to make an awkward turn around $k$. After a reduction proportional to a reduced null degrees of freedom, the posterior mean $E_k(\lambda|x):=w$ increases steadily with $x$, which is what ought to be. Figure \ref{correctedvalue} also shows that one-layer correction is inadequate because its values $u$ are larger than the two-layer values $v$ for all values of $x$ except for small values on the extreme left of the picture.

\subsection{Posterior variance and related results}
The following theorem is the main result of this subsection. It gives an expression of the posterior variance, which makes possible direct estimation from sample data.
\begin{theorem}\label{pv1}
In model (\ref{model}) with $k\geq 2$ and unspecified prior distribution $G(\lambda)$, the posterior variance for $x>0$ equals
\begin{equation}\label{pvar1}
{\rm var}_k(\lambda|x)=4E_{k-4}[4J(J-1)|x]l''_k(x)
+\{E_{k-4}[4J(J-1)|x]-E_{k-2}(2J|x)^2\}(1+2l'_k(x))^2.
\end{equation}
\end{theorem}
Theorem \ref{pv1} has the posterior variance as the sum of two terms. If the marginal likelihood is log concave, the first term is negative because $l''_k(x)$ is negative and its multiplier $E_{k-4}[4J(J-1)|x]$ is positive. The second term, involving $(1+2l'_k(x))^2$, must be positive because the posterior variance would be negative otherwise. It then follows that the multiplier $E_{k-4}[4J(J-1)|x]-E_{k-2}(2J|x)^2$ must be positive. This multiplier roughly corresponds to the variance of $2J$, if we ignore the difference between the second factorial moment and the second moment of $2J$, and the difference between $k-4$ and $k-2$ in the second moment and the first moment of $2J$, respectively. With all these modifications to facilitate interpretation, (\ref{pvar1}) presents the posterior variance of $\lambda$ as the adjusted variance of $2J$ by multiplying $(1+2l'_k(x))^2$, followed by a deduction proportional to the second derivative of marginal log likelihood times the second moment of $2J$.

In (\ref{pvar1}), we know how to estimate each part on the right hand side from the data directly except for $E_{k-4}[4J(J-1)|x]$, which is the task we now undertake.

\begin{lemma}\label{J2lemma}
Under model (\ref{model}), the second factorial moment of $J$ can be expressed in terms of
the marginal densities with different null degrees of freedom via
\[E_{k-4}[4J(J-1)|x]=\frac{x^2g_{k-8}(x)}{g_{k-4}(x)}-\frac{2(k-6)xg_{k-6}(x)}{g_{k-4}(x)}+(k-4)(k-6),\]
where
\[\frac{g_{k-4}(x) }{ g_k(x)} = 4l_k^{''}(x)+[1+2l_k^{'}(x)]^2,\]
\[\frac{g_{k-6}(x) }{ g_k(x)} = 8l_k^{(3)}(x)+12l_k^{''}(x)\{1+2l_k^{'}(x)\}+\{1+2l_k^{'}(x)\}^3,\]
\[\frac{g_{k-8}(x) }{ g_k(x)} = 16l_k^{(4)}(x)+32l_k^{(3)}(x)\{1+2l_k^{'}(x)\}+24l_k^{''}(x)\{1+2l_k^{'}(x)\}^2
+48[ l_k^{''}(x) ]^2+\{1+2l_k^{'}(x)\}^4,\]
where $l_{k}^{(3)}(x)$ and $l_k^{(4)}(x)$ are the third and fourth order derivative of $l_k(x)$, respectively.
\end{lemma}

In Lemma \ref{J2lemma}, we apply Lemma~\ref{Lemma-1} successively, which results in decreasing null degrees of freedom from $k$ to $k -2, k-4, k-6, \cdots$ and increasing order of derivatives for $g_k(x)$. The process begs the question ``how far can it go?". When $k-2i\leq 0$, we lose the natural interpretation of the chi-squared density as the sum of independent standard normal squares. However, the gamma function can be extended to negative non-integer values via analytic continuation. For odd $k$, we can write down $f_{k-2i}(x)$ when $k-2i<0$. Lemma~\ref{Lemma-1} is still valid because the only mathematical property used is $\Gamma(a+1)=a\Gamma(a)$. To be more precise, when $k-2i <0$, $f_{k-2i}$ is no longer a probability density function but we can formally write it down. It would not cause any problem because we never integrate with respect to $x$ in the whole process. Hence, it is straightforward to define $g_{k-2i}$ based on $f_{k-2i}$ for both positive and negative $k-2i$ so long as it is not an even integer.

When $k-2i$ is a nonpositive even integer, we let $f_{k-2i}(x)$ be zero for all $x\geq 0$. This definition preserves the validity of Lemma 1 and thus all subsequent results in our paper for nonpositive even values of $k-2i$. To conclude, $k-2i\leq 0$ would not cause problems in derivation and presentation of formulae such as those in Lemma \ref{J2lemma} so long as we define the corresponding $f_{k-2i}$ as described above.

The following theorem answers affirmatively the existence question of the posterior mean and the posterior variance. The proof is given in the appendix.

\begin{theorem}\label{exist}
Let the prior distribution $G(\lambda)$ of the model (\ref{model}) be any probability distribution over $[0, \infty)$. Let $k\geq 2$, then both the posterior mean and the posterior variance
\[E_k(\lambda|x)<\infty, \;\; {\rm var}_k(\lambda|x)<\infty\]
exist and are finite for each $x>0$.
\end{theorem}

\subsection{Effect size posterior intervals}
Next we will examine the performance of posterior intervals of the form
\begin{equation}\label{pintv}
E_k(\lambda|x)\pm z_{0.95}\sqrt{{\rm var}_k(\lambda|x)},
\end{equation}
where $z_{0.95}=1.645$, the 95\% quantile of the standard normal. Posterior intervals of this form are first introduced by Efron (2010); see (11.67) on p. 229 of Efron (2010). 
\begin{figure}[ht]
\centering
\includegraphics[width=0.8\textwidth]{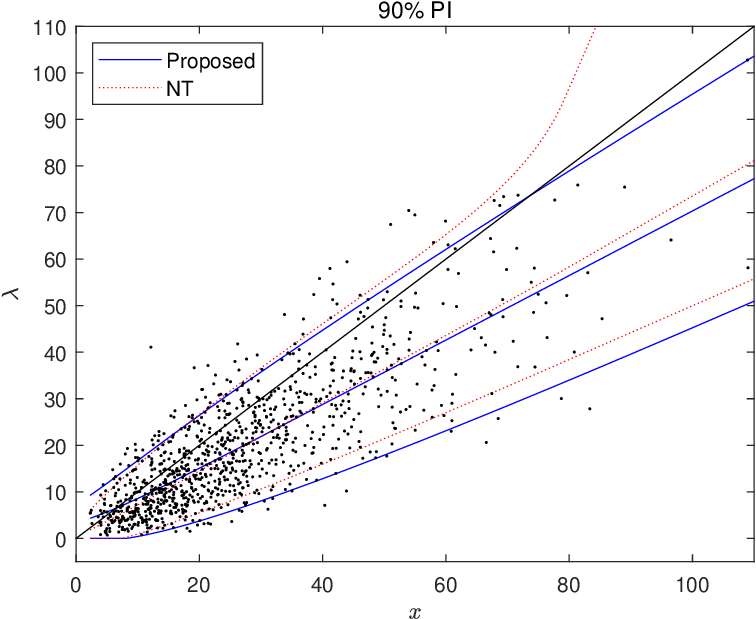}\\
\caption{\textsl{Degrees of freedom $k=7$; the prior density is gamma with shape and scale parameters $\alpha= 2, \beta = 10$, respectively. }}
\label{CI_1}
\vspace{-0.2cm}
\end{figure}

\begin{figure}[ht]
\centering
\vspace{-0.2cm}
\includegraphics[width=0.8\textwidth]{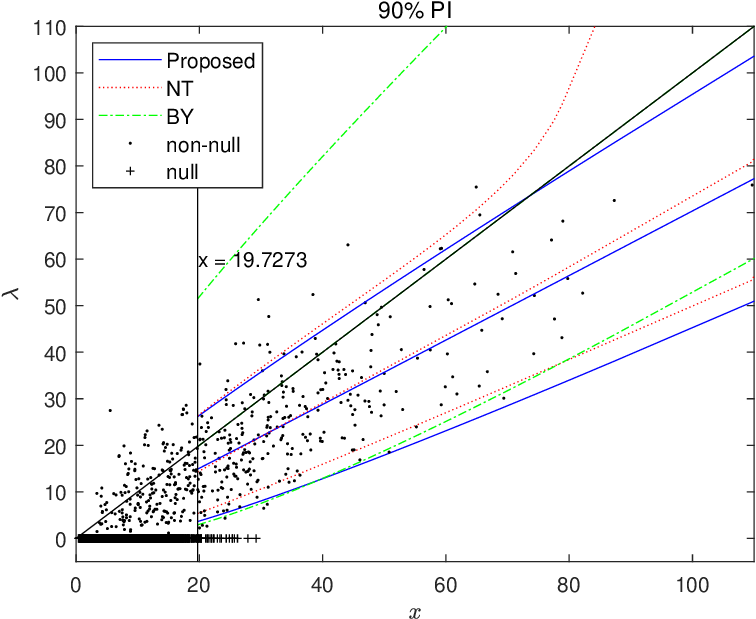}\\
\caption{\textsl{The prior distribution has a point mass at 0, otherwise the same as Figure \ref{CI_1}.}}\label{CI_2}
\vspace{-0.2cm}
\end{figure}

The simulation in Figure \ref{CI_1} is based on $1000$ repetitions. The coverage rates of the proposed method and the normal transformation method (NT) are $90.2\%$ and $87.3\%$, respectively. Figure \ref{CI_2} is obtained from $5000$ repetitions, $90\%$ are sampled from the chi-squared distribution with $k=7$ and 10\% from the same noncentral chi-squared distribution as in Figure~\ref{CI_1}. With ${\rm FDR}=0.1$, the BH procedure selects $314$ cases at cutoff value $=19.7273$, among them $285$ cases are non-null with empirical FDR $=0.0924$. The coverage rates for the proposed, NT's, and BY's methods are $260/285=91.23\%$, $249/285=87.37\%$, and $277/285=97.19\%$, respectively.

Since the prior distribution in Figure \ref{CI_2} has a point mass at zero $\pi_0=P\{\lambda=0\}=0.9$, the formulae for posterior mean and variance in Figure \ref{CI_2} need to be adjusted. Let  ${\rm fdr}_k(x)$ be the {\it local false discovery rate}
\[{\rm fdr}_k(x):=P(\lambda=0 | x)=\frac{\pi_0f_k(x)}{g_k(x)};\]
see Chapter 5 of \cite{Efron:2010}. Then
\begin{eqnarray}\label{pointmass}
E_k(\lambda^i | x)&=& E_k(\lambda^i | x,\lambda=0){\rm fdr_k(x)}+E_k(\lambda^i | x,\lambda>0)(1-{\rm fdr}_k(x)) \nonumber \\
&=& E_k(\lambda^i | x,\lambda>0)(1-{\rm fdr}_k(x)).
\end{eqnarray}
Let $E^1_k(\lambda):= E_k(\lambda | x,\lambda>0)$ and ${\rm var}_k^1(\lambda):={\rm var}_k(\lambda | x,\lambda>0)$.
Applying (\ref{pointmass}) with $i=1, 2$, we obtain the posterior mean and posterior variance in Figure \ref{CI_2}
\begin{equation}\label{H_a}
E^1_k(\lambda)=\frac{ E_k(\lambda | x)}{1-{\rm fdr}_k(x)},\;\; {\rm var}_k^1(\lambda)=\frac{{\rm var}_k(\lambda | x)}{1-{\rm fdr}_k(x)}-{\rm fdr}_k(x)E^1_k(\lambda)^2,
\end{equation}
where $E_k(\lambda | x)$ is given by (\ref{pmean2}), ${\rm var}_k(\lambda | x)$ by Theorem \ref{pv1}; see p. 228 of \cite{Efron:2010}.

\cite{BY:2005} pioneered interval estimates, controlling for the false coverage rate (FCR). In our study, their intervals indeed keep FCR ($=1-277/314=0.1178$) close to the desired level at the cost of overshooting the intended coverage rate. The intervals are much wider than the proposed ones with the lower half covering a lot more points than the upper half, indicating re-center is needed.

 The FCR, however, is unfit for the purpose of the posterior interval. The interpretation of the coverage rate of posterior intervals in Figure \ref{CI_2} is $P(\lambda=0\;|\;x\geq 19.73) \leq 10\%$, otherwise
$P(\lambda \in {\rm posterior\;\;interval}\;|\;x\geq 19.73, \lambda>0)\approx 90\%$.
{\it This bifurcated interpretation is inherited from the model and consistent with our goal to estimate the effect size under the alternative hypothesis}. The BY's interval is designed to control coverage rate under both null and alternatives hypotheses $P(\lambda\in {\rm BY's\;\;interval}\;|\;x\geq 19.73)\approx 90\%$. In other words, BY's method combines two different groups of cases that we want to separate.

In Figures \ref{CI_1} and \ref{CI_2}, the normal transformation method (NT) first turns chi-squared values into z-values via $Z=\Phi^{-1}(F_k(X))$, where $F_k$ is the chi-squared CDF and $X$ is noncentral chi-squared distributed with noncentrality parameter $\lambda$, both of $k$ degrees of freedom. Next find the posterior interval for $\mu=E(Z)$, using Tweedie's formula (\ref{normal}) for the normal distribution, then convert the interval of $\mu$ to the interval of $\lambda$. For various priors, the NT method works well for small $\lambda$ but not so for large $\lambda$. In practice, large $\lambda$ values are more interesting and deserve more attention. The NT method runs into problems due to numerical instability of Gaussian quantile transformation $\Phi^{-1}$ for values close to 1. We resort to extrapolation in Figures \ref{CI_1} and \ref{CI_2}, otherwise the coverage rate would be even lower. The proposed method, however, does not suffer from such problems. Overall, both methods have reasonable coverage rates, close to 90\%, and the latter is somewhat better.

\begin{figure}[ht]
\centering
\begin{tabular}{cc}
\includegraphics[width=0.45\textwidth]{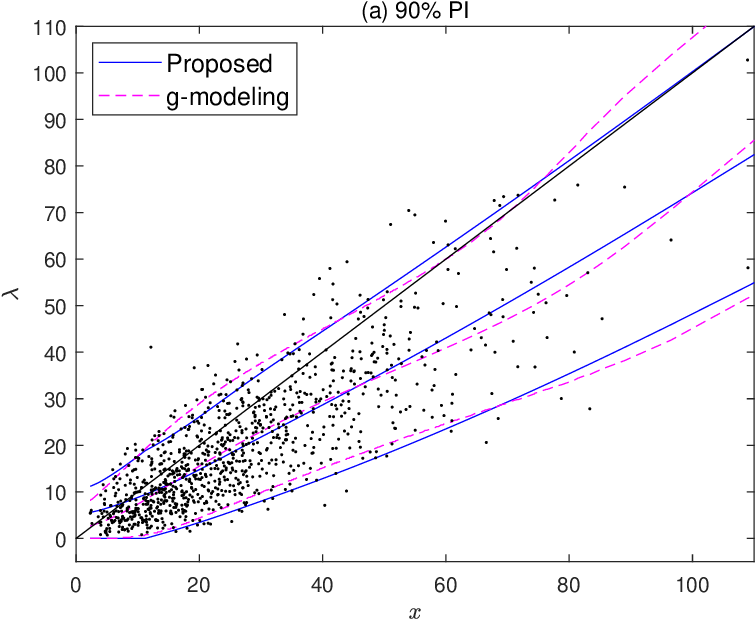}&
\includegraphics[width=0.45\textwidth]{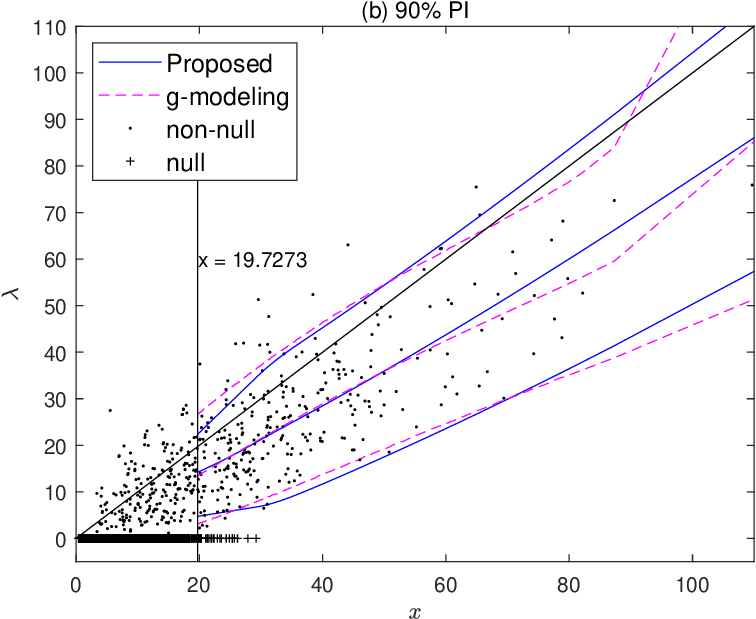}
\end{tabular}
\caption{\textsl{The settings of (a) and (b) are the same as Figure~\ref{CI_1} and Figure~\ref{CI_2}, respectively.
}}
\label{CI_gg}
\vspace{-0.2cm}
\end{figure}
Another way to construct posterior intervals is the $g$-modeling method of \cite{Efron:2016}, which estimates the prior density $g$ via deconvolution; also see \cite{YW:2019}. \cite{Efron:2016} adopted exponential family modelling of $g(\lambda)$. Figure~\ref{CI_gg} compares 90\% posterior intervals of the $g$-modelling method to the proposed method, where the log-density derivatives are estimated by the PLE method in Section~\ref{Sec-3a}. In Figure~\ref{CI_gg}(a), the coverage rates of the proposed method and the $g$-modeling method are $90.4\%$ and $90.2\%$, and 90.2\% and 91.6\% in Figure~\ref{CI_gg} (b), respectively. The coverage rates of both methods are quite close to each other and to the nominal level.

\subsection{Justification for posterior interval}
Figures~\ref{CI_1} and \ref{CI_2} show that the posterior interval has coverage probability close to the nominal level. In both simulation studies, the posterior interval performs as anticipated. We now explain why one can expect the posterior interval for large $x$ values to have coverage probability close to the desired level in general.
By (\ref{bndnc}), we have an upper bound for the non-central chi-squared density
\[f_{k,\lambda}(x)\leq h_k(x)e^{-(\sqrt{\lambda}-\sqrt{x})^2/2},\]
where $h_k(x)$ is a polynomial and inconsequential for it is cancelled in both the numerator and denominator of the posterior density $g_k(\lambda|x)=g_k(\lambda,x)/g_k(x)$. More detailed analysis refines the preceding upper bound
$h_{k,1}(\lambda{x})e^{-(\sqrt{\lambda}-\sqrt{x})^2/2}\leq \frac{f_{k,\lambda}(x)}{h_k(x)}\leq h_{k,2}(\lambda{x})e^{-(\sqrt{\lambda}-\sqrt{x})^2/2}$,
where $h_{k,1}$ and $h_{k,2}$ are two algebraic functions playing minor roles in the shape of the posterior distribution.
As a result, the crucial part of $f_{k,\lambda}(x)$ is the exponential term
\[f^*(\lambda,x)=\exp\{-(\sqrt{\lambda}-\sqrt{x})^2/2\},\]
which is {\it log-concave (thus unimodal) and roughly symmetric} in $\lambda$. If $g(\lambda)$ is log-concave, so is $f^*(\lambda, x)g(\lambda)$. Several well-known families of distributions over $[0,\infty)$ have log-concave densities; e.g. gamma, uniform, truncated normal, truncated logistic, Weibull etc. Furthermore, $\partial^2\log f^*(\lambda, x)/\partial\lambda^2\propto -\sqrt{x}$, indicates that the larger the $x$ the more pivotal $f^*(\lambda, x)$ in the log-concavity (unimodality) of the posterior density.

In conclusion, recognizing $f^*(\lambda, x)$ as the commanding part of noncentral chi-squared density, one can reasonably expect the posterior density $g_k(\lambda|x)\approx f^*(\lambda, x)g(\lambda)/\int_0^\infty f^*(\lambda,x)g(\lambda)d\lambda$ for large $x$ values to be {\it unimodal and roughly symmetric}. This conclusion is supported by numerical and Monte Carlo experiments. Therefore,  the posterior distribution is well summarized by the posterior mean and posterior variance via posterior intervals, just like the mean and variance in prediction intervals for normal distributions.

\section{Log-Density Derivative Estimation}\label{Sec-3a}
To apply the formulae of the posterior mean in Theorem \ref{pm_thm} and posterior variance in Theorem \ref{pv1} to data analysis of latter sections requires the estimation of log density derivatives. Such estimates in Figures~\ref{CI_3}-\ref{triplet} were provided by a penalized least squares (PLE) method introduced in \cite{HV:2005} and \cite{Sasaki_et_al_2014}, which we now describe.

\subsection{Penalized least squares and score matching}
Assume square error loss for log-density derivative estimation
\begin{eqnarray*}
&&\ell(f):=E_k\{f(x)-l'_k(x)\}^2 = \int_0^{\infty}\{f(x)-l'_k(x)\}^2 g_k(x)dx \nonumber \\
&=&\int_0^{\infty}f^2(x)g_k(x)dx-2\int_0^{\infty}f(x)g_k'(x)dx+\int_0^{\infty} (l'_k(x))^2g_k(x)dx \nonumber \\
&=&\int_0^{\infty}f^2(x)g_k(x)dx+2\int_0^{\infty}f'(x)g_k(x)dx-2f(x)g_k(x)\Big{|}_{0}^{\infty}+C,
\end{eqnarray*}
where the last equality follows from {\it integration by parts} and $C$ denotes a constant independent of the estimate $f$. Assume the boundary condition $g_k(0)=g_k(\infty)=0$, then the square error loss is equivalent to
\[\tilde{\ell}(f)=\int_0^{\infty}f^2(x)g_k(x)dx+2\int_0^{\infty}f'(x)g_k(x)dx,\]
which suggests that one should minimize the sample version of square error loss
\begin{equation}\label{loss}
\ell_m(f)=\frac{1}{m}\sum_{i=1}^{m}f^2(x_i)+\frac{2}{m}\sum_{i=0}^{m}f'(x_i).
\end{equation}
The above described technique leading to (\ref{loss}) is referred to as {\it score matching} by \cite{HV:2005}. Suppose that the desired estimate $f(x)$ comes from a space ${\cal S}$ spanned by a collection of basis functions $\{B_i(\cdot), i=1,\cdots, m\}$
\[{\cal S}:=\{f\; \mid \; f(x)=\sum_{i=1}^m \beta_iB_i(x),{\bs \beta}\in \Re^m \},\]
where ${\bs \beta}=(\beta_1,\cdots,\beta_m)^\top$. In Figures~\ref{CI_3}-\ref{triplet}, we adopt B-spline basis functions.

Adding the $\ell^2$-regularizer to the square error loss (\ref{loss}) results in the PLE criterion
\begin{equation}\label{PLE}
\min_{\bs \beta}\left\{\frac{1}{m}\sum_{i=1}^{m}{\bs \beta}^\top{\bf B}(x_i){\bf B}^\top(x_i){\bs \beta}+\frac{2}{m}\sum_{i=1}^{m}{\bs\beta}^\top{\bf B}'(x_i)+\tau{\bs \beta}^\top{\bs \beta}\right\},
\end{equation}
where ${\bf B}(x)=(B_1(x),\cdots, B_m(x))^\top$, and ${\bf B}'=d{\bf B}/dx$. The solution of (\ref{PLE}) equals
\[\hat{\bs \beta}=-\left(m^{-1}\sum_{i=1}^{m}{\bf B}(x_i){\bf B}(x_i)^\top+\tau{\bf I}\right)^{-1}\left( m^{-1}{\sum_{i=1}^{m}\bf B'}(x_i) \right).\]
As usual, the tuning parameter $\tau$ is chosen by cross-validation. Accordingly, the first order log-density
derivative is estimated by $\hat{l}_{k}^{'}(x):= \mathbf{B}^\top(x)\hat{\bs \beta}$. The second derivative can be estimated analogously. Specifically, we adopt the score matching technique to estimate $g_k''(x)/g_k(x)$, and then convert it to an estimate of $l''_k(x)$ via  $\widehat{g''_k/g_k}(x)-\{\hat{l}_{k}^{'}(x)\}^2$.

\subsection{Discussion of the PLE method}
The PLE method described above is quite fitting for the estimation of the posterior mean and variance, and produces satisfactory results for it estimates the log-density derivatives directly. As we know well, a good density estimate does not necessarily produce accurate estimates of its derivatives and it is better to estimate the density derivatives directly.

The {\it score matching} technique is quite flexible and can be used to estimate higher order log-density derivatives, which appear in the posterior variance formula. Even though PLE can handle higher order density derivatives, it is helpful to note that the 3rd and 4th derivatives $l^{(3)}_k(x), l^{(4)}_k(x)$ are essentially zero for large $x$ values. To see this, let's consider
\begin{equation}\label{l_k'}
l_k^{'}(x) = \frac{k/2-1}{x}-\frac{1}{2}+\frac{\int \exp(-\lambda/2) \frac{d}{dx}\delta_k(\lambda x)dG(\lambda)}{\int \exp(-\lambda/2)\delta_k(\lambda x)dG(\lambda)},
\end{equation}
where \[\delta_k(\rho)\equiv\sum_{j=0}^{\infty}\frac{(\rho/4)^j}{j!\Gamma{(k/2+j)}}.\]
Differentiating $l'_k(x)$ to get 3rd and 4th derivatives of the log-density reveals that the first terms are
$O(x^{-3})$ and $O(x^{-4})$, respectively, which converge to zero very fast as $x\rightarrow\infty$. More elaborated analysis of the last term shows that the convergence rates are actually $O(x^{-5/2})$ and $O(x^{-7/2})$.

In brief, estimating log-density derivatives directly without first estimating lower order derivatives, the PLE method suits our purpose here. The higher order derivatives in posterior variance are negligible for large $x$ values. These facts together adequately  address the estimation issue of log-density derivatives.

\section{Interpretation of Posterior Mean Estimates}\label{Sec-4}
In Theorems \ref{pm_thm} and \ref{pv1}, after plugging in the estimated log density derivatives, we obtain estimates for the posterior mean and variance of noncentrality parameter. Here we suggest ideas to interpret such estimates. 
The ideas are grouped under two definitions: {\it posterior significance} for point estimates and {\it posterior dominance} for interval estimates.
\subsection{Posterior significance}

\noindent{\it Begin with average per DF}\\
The noncentrality parameter $\lambda=\sum_1^k\mu_i^2$ equals the sum of squared effects over $k$ components. Thus the posterior mean $E_k(\lambda|x)$ represents the total contribution from $k$ degrees of freedom (DF). Having observed $X=x$ without information on individual contribution from each DF, it is natural to begin our analysis with the average $E_k(\lambda|x)/k$ per DF. Average per DF is also the contribution allocated to each DF by the maximum entropy distribution. By the maximum entropy principle, after observing the total contribution, the best guess of the state is that the individual contribution from each DF equals the average per DF because such a state has maximum entropy. Thus for interpreting a point estimate of the posterior mean $E_k(\lambda|x)$, it makes sense to base that on the average per DF. \\

\noindent{\it Calibrate average per DF}\\
By (\ref{model}), the observed value $x$ has chi-squared distribution with $(k+2J)$ DF
\[\underbrace{Z^2_1+\cdots+Z^2_k}_{\rm noise}+\underbrace{Z^2_{k+1}+\cdots+Z^2_{k+2J}}_{\rm signal}=x, \]
where $Z_i$'s are independent standard normal random variables. Under the null hypothesis, $x$ has chi-squared distribution with $k$ DF so the sum of the first $k$ $Z^2_i$'s contains no information on $\lambda$. If its value were known, one would subtract that from $x$, then the sum of remaining $2J$ $Z^2_i$'s is an unbiased estimate of $\lambda$. Thus we can rightfully refer to the sum of the first $k$ $Z^2_i$ as `noise' and the sum of remaining $2J$ $Z^2_i$'s as `signal'. Both the signal and noise consist of chi-squared one random variables as the basic unit. In particular, the contribution to $x$ per DF under the null hypothesis is chi-squared one distributed. Therefore we calibrate the average per DF by chi-squared one. These considerations call for the following definition.

\begin{description}
\item {\bf Definition 1} $E_k(\lambda|x)$ is posteriorly significant at $\alpha$-level per degrees of freedom, if
\[\frac{E_k(\lambda|x)}{k}\geq z_{1-\alpha/2}^2,\]
\end{description}
where $z_a$ is the $a$-quantile of a standard normal distribution so that $P(\chi_1^2\geq z_{1-\alpha/2}^2)=\alpha$.
For example, $E_k(\lambda|x)$ is posteriorly significant at $10\%$ level with $k=7$ DF, it means
\[\frac{E_k(\lambda|x)}{7}\geq z_{0.95}^2=(1.645)^2 \Leftrightarrow E_k(\lambda|x)\geq 18.9.\]
 If a posterior mean is posteriorly significant at 10\%, then the average per DF is larger than 90\% of the noises. Knowing that the noise and selection bias have already been removed in the posterior mean supposedly, it is very significant that a point estimate of the posterior mean exceeds 90\% noises per DF and such a result should be highlighted in data analysis.

\subsection{Posterior dominance}
\noindent{\it Interval estimate}\\
As interval estimates accommodate sampling variation, posterior intervals accommodate variation when sampling from the posterior distribution. The posterior interval of (\ref{pintv}) gives the range that covers about 90\% of the values from the posterior distribution at $x$. \\

\noindent{\it A point estimate above an interval estimate}\\
If a posterior mean is above another 90\% posterior interval, the former posterior mean not only is larger than the latter posterior mean but also larger than 90\% of the values sampled from the latter posterior distribution. This type of strong results leads to

\begin{description}
\item {\bf Definition 2} $E_k(\lambda|x)$ {\it posteriorly dominates} $E_k(\lambda|x')$ at $100(1-\alpha)\%$ level , if
\[E_k(\lambda|x)\geq E_k(\lambda|x')+z_{1-\alpha/2}\sqrt{{\rm var}_k(\lambda|x')}.\]
\end{description}

For instance,
\[E_k(\lambda|x)\geq E_k(\lambda|x')+z_{0.95}\sqrt{{\rm var}_k(\lambda|x')}
= E_k(\lambda|x')+1.645\sqrt{{\rm var}_k(\lambda|x')},\]
then $E_k(\lambda|x)$ dominates $E_k(\lambda|x')$ with posterior probability at least 90\%. The `at least' here is because $E_k(\lambda|x)$ is at least as high as the upper limit of the interval at $x'$ and thus the concerned posterior probability is actually close to $95\%$. \\

\noindent{\it One interval estimate above another}\\
The strongest possible type of results considered in this section is when the posterior interval at $x$ is entirely above that at the other value $x'$. It indicates that except for a small bottom portion, the former posterior distribution has all its values larger than all values except for a small top portion of the latter posterior distribution.

\begin{description}
\item {\bf Definition 3} The posterior interval at $x$ {\it dominates} the other at $x'$ at $100(1-\alpha)\%$ level, when
\[E_k(\lambda|x)-z_{1-\alpha/2}\sqrt{{\rm var}_k(\lambda|x)}\geq
{E}_k(\lambda|x')+z_{1-\alpha/2}\sqrt{{\rm var}_k(\lambda|x')}.\]
\end{description}
In the next two sections, we will apply the concepts of {\it posterior significance} and {\it posterior (interval) dominance} defined above to interpret the results in the simulation study and real data examples, where the posterior mean and intervals are estimated from the sample data.

\section{Variable Selection via Chi-squared Statistic}\label{Sec-5}
Let $X_1,\cdots, X_{100}$ be i.i.d. Bernoulli distributed with $P(X_i=0)=P(X_i=1)=1/2$. Consider the following model for the response variable $Y$,
\begin{equation}\label{XOR}
Y=\left\{
\begin{array}{lc}
X_1+X_2-2X_1X_2+\epsilon & {\rm prob.\;\; 0.5}; \\
X_3+X_4+X_5-2(X_3X_4+X_3X_5+X_4X_5)+4X_3X_4X_5+\epsilon & {\rm prob.\;\; 0.5},
\end{array}
\right.
\end{equation}
where $\epsilon$ is independent normal distributed with mean zero and standard deviation $0.5$. The expression $X_1+X_2-2X_1X_2$ in (\ref{XOR}) is the celebrated XOR function, a classical example in neural networks demonstrating that a nonlinear function can be learned via hidden layers, whereas the part involving $\{X_3,X_4,X_5\}$ is the triplet version of XOR; see e.g. \S 6.1 of \cite{GBC:2017}. The model (\ref{XOR}) has nature flip a fair coin. If head (tail) appears, then $Y$ equals the XOR function (the triplet version) plus the random error $\epsilon$, respectively.
\begin{figure}[ht]
\centering
\begin{tabular}{cc}
\includegraphics[width=0.45\textwidth]{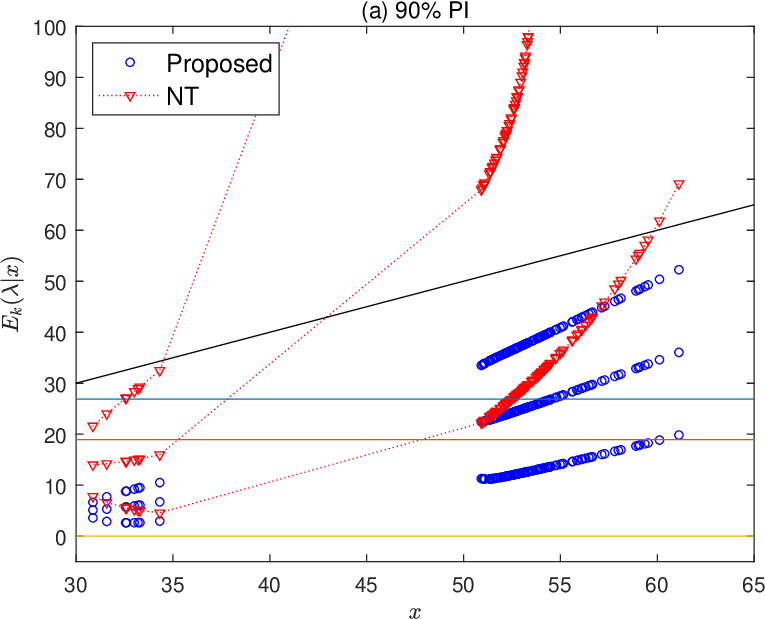} &
\includegraphics[width=0.45\textwidth]{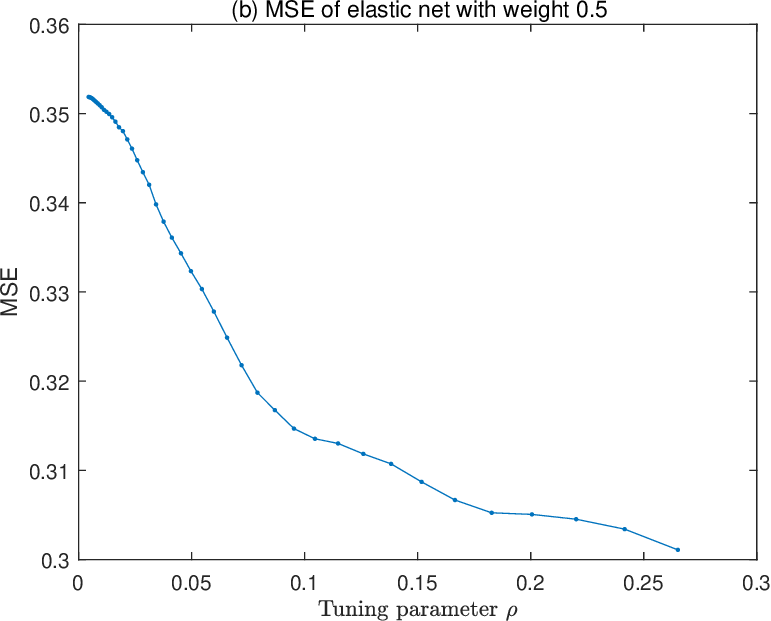} \\
\end{tabular}
\caption{\textsl{(a) The circle-marker curves are the estimated upper limit, posterior mean, and lower limit of the posterior interval in (\ref{pintv}), respectively. The triangle curves are for the NT method. (b) The mean squared error (MSE) is based on $10$-fold cross-validation using the elastic net that equally combines the $\ell_1$ and $\ell_2$ penalties.
} }\label{CI_3}
\end{figure}

Simulate 300 i.i.d. copies of $(Y,X_1,X_2,\cdots,X_{100})$, where $Y$ obeys (\ref{XOR}), as the data matrix ($300\times 101$). Each triplet $\{X_{i_1},X_{i_2},X_{i_3}\}$ divides the data into 8 groups according to $(X_{i_1},X_{i_2},X_{i_3})=(0,0,0),(0,0,1),\cdots,(1,1,1)$. Let $\bar{Y}$ be the sample mean of $Y$'s and $\bar{Y}_j$ be the mean of $n_j$ observations in $j$-th group and $\hat{\sigma}^2$ be the sample variance of $Y$'s.

To see how the variables in a triplet {\it jointly} affect the value of $Y$, we employ the chi-squared statistic
\[Q\{X_{i_1},X_{i_2},X_{i_3}\}=\sum_{j\in {\cal P}_i}\frac{n_j(\bar{Y}-\bar{Y}_j)^2}{\hat{\sigma}^2},\]
where ${\cal P}_i$ represents the partition of 8 groups by $i$-th triplet. Thus $Q$ equals the sum of squared standardized deviations of group means from the sample mean over 8 groups. When $X_1,\cdots, X_5 \notin \{X_{i_1},X_{i_2},X_{i_3}\}$, that is, the triplet contains no causal variables, $Q$ is asymptotically chi-squared distributed with $k=8-1=7$ degrees of freedom since the population mean estimated by the sample mean costs one degree of freedom. 

We compute the $Q$-values for all $100\times99\times98/6$ triplets. Then apply the BH procedure with ${\rm FDR}=10\%$ to select $107$ triplets. The results are given in Figure \ref{CI_3}(a). The far right point concerns the $Q$-value of $\{X_3, X_4, X_5\}$; the middle trunk consists of $Q$-values of $98$ triplets $\{X_1, X_2, X_i\}$, $i=3, \ldots, 100$; the horizontal lines are $y=0$, $y=1.645^2\times k$, and $y=1.96^2\times k$ with $k=7$, respectively.

The triplets $\{X_3, X_4,X_5\}$ and $\{X_1,X_2,X_i\}, i=3,4,\cdots, 100$ are distinctively separated from other triplets. These 99 triplets have their effect size estimates attain {\it posterior significance} at 10\% level. They also achieve {\it posterior interval dominance} over other triples at 90\% level. That is, the lower limits of 99 posterior intervals on the right all exceed the upper limits of the other 8 posterior intervals on the left. Therefore, the data provide strong, if not overwhelming, evidence that these 99 triplets contain causal variables that influence the values of $Y$. The preceding analysis suggests that ${\rm FDR}=8/107=7.5\%<10\%$.

 Assume that the underlying model is {\it sparse}, then one can infer that $X_i, i=3,\cdots, 100$ are included in the high-scored triplet $\{X_1,X_2,X_i\}$ because they are free-riders with $\{X_1,X_2\}$. Dropping $X_i$ from $\{X_1,X_2,X_i\}$ yields a more significant result for $\{X_1,X_2\}$. Therefore, Figure \ref{CI_3}(a) indicates that the proposed method can correctly identify two signaling modules, $\{X_1, X_2\}$ and $\{X_3,X_4,X_5\}$ from many irrelevant $X$-variables.

The posterior mean estimated by the normal transformation method for the top 99 triplets are larger and appear more significant. However, these estimates are unreliable for two reasons. First, the corresponding posterior intervals are too wide. More importantly, these posterior mean estimates do not carry shrinkage effect and are larger than the corresponding $Q$-values. This is not supposed to happen as the estimates are corrected for selection bias. These results occur possibly due to unsatisfactory density derivative estimates by Poisson regression method that comes with the normal transformation method.

For comparison, we expand the design matrix by adding all of the two-way and three-way interaction terms and run the LASSO analysis of  \cite{Tibshirani:1996} to select the relevant variables. We adopt $10$-fold cross-validation to estimate the mean squared error (MSE) curve for given values of tuning parameter $\rho$ in Figure~\ref{CI_3}(b).

Persistently decreasing MSE implies that the LASSO fails to select any variable, possibly because it is designed to detect main effects instead of interaction. The LASSO is correlation-based and under (\ref{XOR}) all relevant variables $X_1,\cdots, X_5$ have zero correlation with $Y$ so that it is difficult for the LASSO to pick up the signal in these variables. Consequently, other correlation-based variable selection methods are likely to experience the same difficulty with (\ref{XOR}). Further discussion are in Section \ref{Sec-7}.

\section{Real Data Examples}\label{Sec-6}
\subsection{Differences in gene expression among ethnic groups}
We apply the proposed method to the microarray data first analyzed in \cite{spielman_etal_2007} to characterize genetic variation among four ethnic groups. The dataset consists of $p=8793$ annotated genes expression levels of $n=166$ individuals from four populations. These include $60$ European-derived individuals from the Utah pedigrees of the Centre d'Etude
du Polyporphisme Humain (CEU), $41$ Han Chinese in Beijing (CHB), $41$ Japanese in Tokyo (JPT), and $24$ from the Han Chinese in Los Angeles (CHLA). While \cite{spielman_etal_2007} reported that $25\%$ of gene expression levels differ significantly between populations, possibly due to allele frequency differences at cis-linked regulators, some follow-up studies \citep{Leek_etal_2010} criticized the pervasive significance because the populations and processing dates are highly correlated. Strong batch effects such as the processing dates should be accounted for before any significance test is conducted.

For each gene, the following statistical model decomposes the expression level into contributions from three sources: ethnic group membership, latent common factors, and random errors. Specifically, the expression level of $i$th-gene,
\begin{equation*}
\mathbf{Y}_{i} = \mathbf{X}\boldsymbol{\beta}_i+\mathbf{Z}\boldsymbol{\gamma}_i+\bs{\varepsilon}_i,
\end{equation*}
where $\mathbf{X} \in \mathbb{R}^{n \times 4}$ is the contrast matrix for the group membership, $\mathbf{Z}$ denotes the latent common factors, and $\bs{\varepsilon}_i$ is the random error with constant variance $\sigma_i^2$. The latent factor part is estimated by $\hat{\mathbf{Z}}\hat{\boldsymbol{\gamma}_i}$ using the restricted principle component analysis algorithm proposed by \cite{Du_Zhang_2017}. Then a chi-squared statistic is employed to test homogeneity among the four groups. The test statistic $T_i$, which has chi-squared distribution with $k=3$ degrees of freedom asymptotically under the null hypothesis, is computed as
\begin{eqnarray*}
\hat{\bs{\beta}}_i &=& \{\mathbf{X}^\top\mathbf{X}\}^{-1}\mathbf{X}^\top\{\mathbf{Y}_{i}-\hat{\mathbf{Z}}\hat{\boldsymbol{\gamma}}_i\},  \\
T_i &=& \{\mathbf{A}\hat{\bs{\beta}}_i\}^\top
\{\mathbf{A}(\mathbf{X}^\top\mathbf{X})^{-1}\mathbf{A}^\top \}^{-1}
\{\mathbf{A}\hat{\bs{\beta}}_i\}/\hat{\sigma}_i^2,
\end{eqnarray*}
where $\hat{\sigma}_i$ is the root mean squared error of the $i$-th regression and $\mathbf{A} \in \mathbb{R}^{3 \times 4}$ is the dummy matrix for testing the homogeneity hypothesis $\beta_{i2}=\beta_{i3}=\beta_{i4}=0$.

\begin{figure}[ht]
\centering
\includegraphics[width=0.7\textwidth]{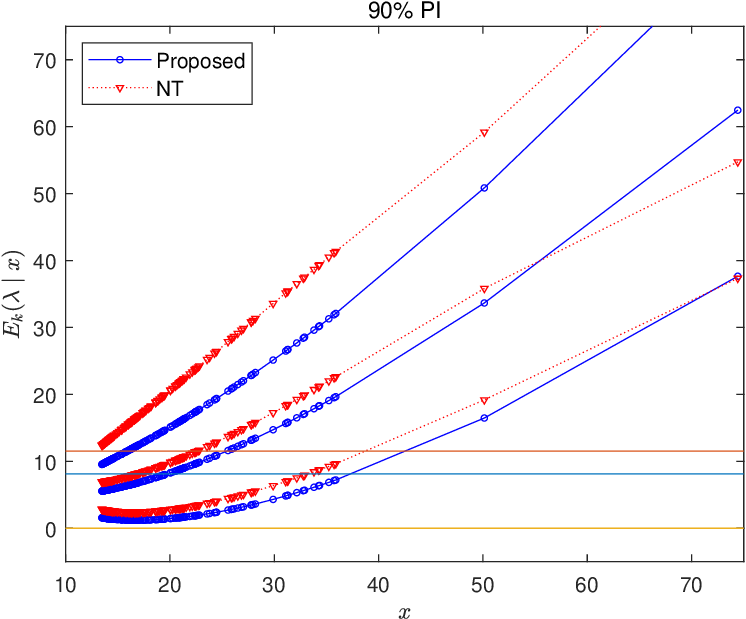}\\
\caption{\textsl{The 90\% posterior intervals for the selected $164$ genes;
the horizontal lines are $y=0$, $y=1.645^2\times k$, and $y=1.96^2\times k$ with degrees of freedom $k=3$, respectively.} }\label{Data_1}
\end{figure}

At FDR = 0.10, 164 genes were identified and their corresponding posterior intervals are shown in Figure~\ref{Data_1}.
The results indicate that 58 and 26 out of 164 genes achieve {\it posterior significance} at 10\% and 5\% levels, respectively. The two genes on the far right with the largest effect sizes attain {\it posterior dominance} over other genes at 90\% level.

The $90\%$ posterior intervals of both the proposed and the normal transformation methods are all above zero, but the latter intervals are wider and further away from zero, with 87 and 33 genes posterior significant at 10\% and 5\%, respectively. These results are consistent with findings from simulation studies that the normal transformation method has a smaller shrinkage effect than the proposed method for large test statistic values.

\subsection{Gene expression profiling and breast cancer metastasis}
The data set of this example is taken from the study in \cite{van't Veer:2002}. About $30\%$ of breast cancer patients would benefit from chemotherapy or hormonal therapy, which reduces the risk of distant metastasis. The other $70$-$80\%$ of patients would survive without the adjuvant therapies. \cite{van't Veer:2002} advocated using gene expression profiling to select breast cancer patients for such therapies. The original data contain the expression levels of over 20,000 genes for each of 78 breast cancer patients, which they used as the training set in developing their prognosis classifier. After pre-processing, 4918 gene expression levels are retained for our analysis.

The paper, \cite{van't Veer:2002}, is well cited and the dataset has been extensively analyzed by numerous authors using a wide variety of feature selection and classification methods; see e.g. \cite{WLZH:2012} for a brief review. The classification error rates (with correct cross-validation) reported in the literature up to year 2011 typically are around 30\%. Some authors suspect that the high error rates are due to interactions among genes have not been accounted for and methods applied to the dataset only consider the main effect of genes. \cite{WLZH:2012} proposed an interaction-based feature selection and classification method, which yields a cross-validated error rate of 8\%. However, it focused on prediction and statistical inference was largely ignored. Here, under model (\ref{model}), we would like to see if there is statistical evidence of higher order interactions in the data.

The response variable $Y=0$, if the patient is free from disease in an interval of at least five years after initial diagnosis, and $Y=1$, if the patient develops distant metastasis in five years. Order expression levels of each gene for 78 patients from large to small. Let $X_i=1$ (high expression) or $X_i=0$ (low expression) depending on the patient belong to the upper or lower half, respectively, in the ordered gene expression levels.

We examine the interaction among 4918 genes via the chi-squared statistic
\[Q(X_{i_1},X_{i_2},X_{i_3})=\sum_{j\in {\cal P}_i}\frac{n_j(\bar{Y}-\bar{Y}_j)^2}{\hat{\sigma}^2},\]
where ${\cal P}_i$ represents the partition of the sample into 8 groups by the triplet $(X_{i_1},X_{i_2},X_{i_3})=(0,0,0),(0,0,1),\cdots,(1,1,1)$. The sample mean, group means, and the sample variance of $Y$ are denoted by $\bar{Y}, \bar{Y}_j$, and $\hat{\sigma}^2$, respectively. We use the statistic $Q$ to capture three-way interactions in triplets. The reason for focusing on three-way interaction is that two-way interaction is deemed too simple to explain complex disease such as breast cancer, while four-way interaction is computationally prohibiting.

\begin{figure}[ht]
\centering
\includegraphics[width=0.7\textwidth]{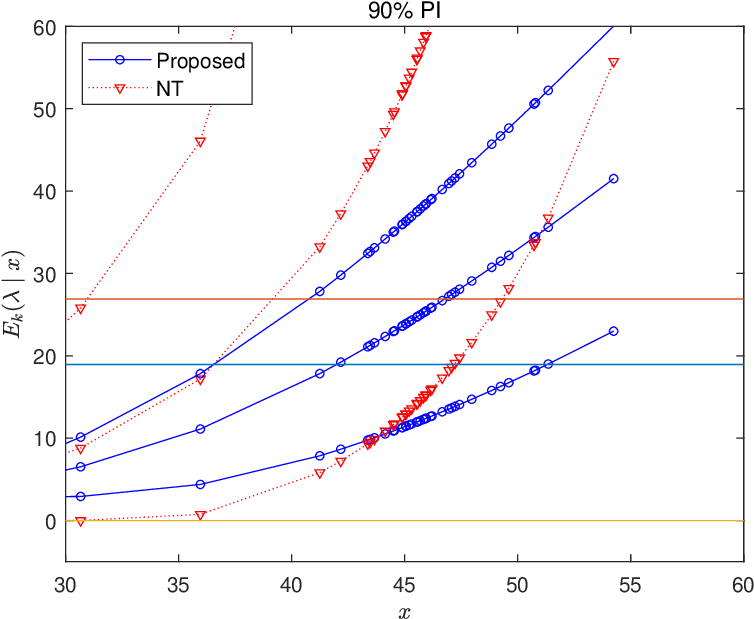}\\
\caption{\textsl{The 90\% posterior intervals for 37 triplets. Two triplets on the far left are from 10,000 randomly selected ones. The other 35 are from the top 50,000 triplets. The horizontal lines are $y=0$, $y=1.645^2\times k$, and $y=1.96^2 \times k$, with $k=7$, respectively.} }\label{triplet}
\end{figure}

We compute the chi-squared statistic $Q$ for all $4918\times4917\times4916/6$ triplets. It takes about 28 hours on a PC with two CPU of 2.66G Hz each. Thus with parallel computing, it can be done in a few hours, a very manageable task. From the top 50,000 $Q$-values, we identify 35 non-overlap triplets by going down the ordered list, removing all triplets overlapped with previously retained ones of higher $Q$-values. This is for reducing the dependence among overlapping triplets. Then randomly select 10,000 triplets and combine with the 35 non-overlapped ones to produce the posterior bands in Figure \ref{triplet}.

With a little over 10,000 triplets, we can cover nearly all genes (missing only 7 out of 4918 genes in a simulation experiment). It is reasonably close to the minimum number of randomly selected triplets that achieves nearly complete coverage (a naive approximation via coupon collector's problem yields an expected value of around 12,000). If we go well beyond 10,000 triplets, then there would be too many overlapped genes and the dependence problem becomes severe. On the other hand, using substantially below 10,000 triplets not only leaves a sizable subset of genes uncovered but also produces unreasonable posterior mean estimates. 
Therefore, with a random sample of around 10,000 triplets, we have a comprehensive coverage of the whole gene pool and not too many overlapped genes in triplets, which may otherwise distort the findings due to dependence.

Using the BH procedure with FDR = 0.10, 37 triplets are selected and their corresponding point and interval estimates are in Figure \ref{triplet}. Nearly all (34 out 35) triplets from the top 50,000 triplets attain 10\% {\it posterior significance}, while 11 reach 5\% level. Moreover, 34 triplets on the right achieve {\it posterior dominance} over two triplets on the far left at 90\% level. These results imply that ${\rm FDR}=3/37=8.1\%<10\%$. The preceding analysis indicates that there is considerable evidence of higher order interactions among top-scored triplets. To predict the metastasis status of a breast cancer patient, it is advisable to incorporate higher order interactions as in \cite{WLZH:2012}.

For this dataset, the normal transformation method produces unreasonable results. Specifically, the posterior mean $E_k(\lambda|x)$ for $x>45$ is even larger than $x$, losing the shrinkage effect expected of any selection bias corrected estimate. Furthermore, the posterior intervals are much wider than those of the proposed method. This is particularly so for large $x$ values. These results confirm previous findings from the simulation studies in Section \ref{Sec-3}.2 that the normal transformation method does not perform well for large $x$ values.

Gene-gene interaction or epistasis is much more challenging to analyse than a single gene because of combinatorial explosion. There are just too many gene combinations to investigate. Thus in the literature, many gene-interaction studies confine themselves within established gene sets or genetic pathways; see e.g. \cite{ZLD:2009}. Others studies investigate gene combinations up to certain level (mostly pairwise) and require extremely small p-values to claim significance; see e.g. \cite{Chu:2014}. However, small p-values do not necessarily lead to sizeable effects. The tools developed in this paper make possible the assessment of the effect size of gene interaction without involving biological information. It is interesting to compare the findings with those from existing gene networks/pathways, an issue of quite different nature and we shall not pursue that here.

\section{Concluding Remarks and Discussions}\label{Sec-7}
For large scale inference, controlling FDR is now a standard practice after performing a large number of hypothesis tests. Suppose that controlling FDR selects a small subset of supposedly non-null cases to estimate their effect sizes. In this regards, \cite{Efron:2010} proposed an empirical Bayes method based on Tweedie's formula for normal data.

Tweedie's formula captures selection bias in a simple relationship involving the marginal likelihood, which allows direct estimation by sample data. Here we develop a parallel formulation for chi-squared data. A Bayesian hierarchical model is introduced as the starting point. We also examine a few new phenomena in the resulting Tweedie's formula, which may inspire the construction of new data-dependent procedures for bias correction.

One application of \cite{Efron:2011} is the $t$-test whereas ours is the chi-squared test. Both are time-honored statistical tests. In variable selection for high dimensional data, $t$-tests are useful in linear main effect models, while chi-squared tests can be applied to models with sparsity, nonlinearity, interaction, and mixture features such as the one in Section \ref{Sec-5}.

The simulation study in Section \ref{Sec-5} is intriguing in three aspects. First, having its root in deep learning, (\ref{XOR}) differs from the usual regression models for high dimensional data. It is challenging to identify the causal variables not only because (\ref{XOR}) is highly nonlinear with interactions as large as main effects in size and opposite in sign, but also because all causal variables in a signaling module have to be identified together. An incomplete module behaves like noises since each causal variable has zero correlation with the response variable. Secondly, (\ref{XOR}) bears no resemblance to (\ref{model}). However, statistical tools derived from (\ref{model}) can identify two signaling modules perfectly as if the tools were designed specifically for (\ref{XOR}). Thirdly, the shape of posterior bands in Figures \ref{CI_3}(a) and \ref{triplet} are similar, though the former is based on simulated data and the latter comes from real data. Does it mean that the real data are generated by a mechanism sharing something in common with (\ref{XOR})?

Two real data applications are considered: interaction-based variable selection for high dimensional data paving the way for prediction of breast cancer metastasis, and testing genetic homogeneity among ethnic groups with confounding latent factors. We provide point and interval estimates corrected for selection bias and interpret these estimates to assess different levels of evidence among selected chi-squared statistics. 

 Several issues remain to be explored. We argue in Section \ref{Sec-3}.5 that one can expect the posterior distribution to be unimodal for large $x$. What is the best condition to ensure unimodality of the posterior distribution under (\ref{model})? Unimodality provides support for the posterior intervals proposed in Section \ref{Sec-3}.2. How to estimate and adjust for skewness in the posterior distribution? Skewness adjustment would improve the coverage rate of posterior intervals. We redesign the PLE method in \cite{HV:2005} and \cite{Sasaki_et_al_2014} to provide reliable log-density derivative estimates. What optimal properties does the PLE method have?  All these issues will be addressed in another paper.  
\\
\\
\centerline{\bf\Large Appendix}
\setcounter{section}{1}
\renewcommand{\theequation}{\Alph{section}.\arabic{equation}}
\noindent{\bf Proofs of (\ref{e2j}) and Theorem \ref{pm_thm}}\;\;
Let $p_j\equiv P(J=j)=\int_0^{\infty}P(J=j|\lambda)dG(\lambda)=\int_0^{\infty}e^{-\lambda/2}\frac{(\lambda/2)^j}{j!}dG(\lambda)$ so that $g_k(x)=\sum_{j=0}^{\infty}p_jf_{k+2j}(x)$.
The chi-squared density assures that $xf_{k-4+2j}(x)=(k-4+2j)f_{k-2+2j}(x)$ and
\[E_{k-2}(k-4+2J|x)=\frac{\sum_{j=0}^{\infty}(k-4+2j)p_jf_{k-2+2j}(x)}{g_{k-2}(x)}=\frac{x\sum_{j=0}^{\infty}p_jf_{k-4+2j}(x)}{g_{k-2}(x)}=\frac{xg_{k-4}(x)}{g_{k-2}(x)}.\]
Consequently,
\begin{eqnarray*}
E_{k-2}(2J|x)&=&\frac{xg_{k-4}(x)}{g_{k-2}(x)}-(k-4)=\frac{x[2g'_{k-2}(x)+g_{k-2}(x)]}{g_{k-2}(x)}-(k-4), \\
&=&\frac{2x[2g''_k(x)+g'_k(x)]}{2g'_k(x)+g_k(x)}+(x-k+4) = 2x[\frac{2l_k^{''}(x) }{1+2l_k^{'}(x)}+l_k^{'}(x)]+(x-k+4),
\end{eqnarray*}
where the second equation is obtained by repeatedly applying Lemma~\ref{Lemma-1}. Substituting the preceding equation, which is exactly (\ref{e2j}), into (\ref{pmean1}), we obtain Theorem \ref{pm_thm} and (\ref{pmean2}). \hfill$\Box$
\\
\noindent{\bf Proof of Theorem \ref{pv1}}\;\;
We begin with
\begin{eqnarray}\label{2m}
E_k(\lambda^2|x)&=&\frac{\int_0^\infty\lambda^2\sum_{j=0}^{\infty}e^{-\lambda/2}\frac{(\lambda/2)^j}{j!}f_{k+2j}(x)dG(\lambda)}{g_k(x)}\nonumber \\
&=&\frac{\int_0^\infty4\sum_{j=0}^{\infty}(j+2)(j+1)e^{-\lambda/2}\frac{(\lambda/2)^{j+2}}{(j+2)!}f_{k+2j}(x)dG(\lambda)}{g_k(x)}\nonumber \\
&=&\frac{\int_0^\infty4\sum_{j=0}^{\infty}j(j-1)e^{-\lambda/2}\frac{(\lambda/2)^{j}}{(j)!}f_{k-4+2j}(x)dG(\lambda)}{g_k(x)}
=\frac{E_{k-4}[4J(J-1)|x]}{g_k(x)/g_{k-4}(x)}
\end{eqnarray}
Apply Lemma~\ref{Lemma-1} twice, first on $g_{k-4}$ then on $g_{k-2}$, to obtain
\[\frac{g_{k-4}(x)}{g_k(x)}=\frac{4g''_k(x)}{g_k(x)}+\frac{4g'_k(x)}{g_k(x)}+1,\]
then by (\ref{2m})
\[
{\rm var}_k(\lambda|x)=E_{k-4}[4J(J-1)|x]\left(\frac{4g''_k(x)}{g_k(x)}+\frac{4g'_k(x)}{g_k(x)}+1\right)-E_{k-2}(2J|x)^2\left(1+2\frac{g'_k(x)}{g_k(x)}\right)^2.
\]
In view of
\[l''_k(x)=\frac{g''_k(x)}{g_k(x)}+\frac{g'_k(x)}{g_k(x)}+\frac{1}{4}-\left(\frac{1}{2}+\frac{g'_k(x)}{g_k(x)}\right)^2,\]
the proof is complete. \hfill$\Box$
\\
\noindent{\bf Proof of Lemma \ref{J2lemma}}\;\;
The chi-squared density satisfies
\[f_{k-8+2j}(x)=x^{-2}(k-8+2j)(k-6+2j)f_{k-4+2j}(x).\]
Thus
\begin{eqnarray}\label{j2begin}
E_{k-4}[(k-8+2J)(k-6+2J)|x]
&=&\frac{\sum_{j=0}^{\infty}(k-8+2j)(k-6+2j)p_jf_{k-4+2j}(x)}{g_{k-4}(x)} \nonumber \\
&=&\frac{x^2\sum_{j=0}^{\infty}p_jf_{k-8+2j}(x)}{g_{k-4}}=\frac{x^2g_{k-8}(x)}{g_{k-4}(x)}
\end{eqnarray}
Since
\[E_{k-4}[4J(J-1)|x]=E_{k-4}[(k-8+2J)(k-6+2J)|x]-(k-6)E_{k-4}(4J-2|x)-(k-6)^2,\]
combining (\ref{j2begin}) and (\ref{e2j}) with $k-2$ replaced by $k-4$, we obtain the desired expression for $E_{k-4}[J(J-1)|x]$.
Furthermore, the equations for $g_{k-4}$ and $g_{k-8}$ follow from applying Lemma~\ref{Lemma-1} successively on subscript $k$, which goes backward at step size of 2. \hfill $\Box$
\\
\noindent{\bf Proof of Theorem \ref{exist}}\;\;
We first bound the noncentral chi squared density as follows.
\begin{eqnarray}\label{bndnc}
f_{k,\lambda}(x) &=& \sum_{j=0}^{\infty}\exp(-\frac{\lambda}{2})
\frac{ (\frac{\lambda}{2})^{j} }{j!}f_{k+2j}(x)
=\sum_{j=0}^{\infty}\exp(-\frac{\lambda}{2})
\frac{(\frac{\lambda}{2})^{j} }{j!} \times
\frac{x^{k/2+j-1}}{ \Gamma(k/2+j)2^{k/2+j} }\exp(-\frac{x}{2}) \cr
&=& \frac{x^{k/2-1}}{\Gamma(k/2)2^{k/2}}\exp\{-\frac{\lambda+x}{2}\}
\sum_{j=0}^{\infty}\frac{(\lambda x)^j}{2^{2j}j!(k/2+j-1)\times \cdots \times(k/2)} \cr
&=& \frac{x^{k/2-1}}{\Gamma(k/2)2^{k/2}}\exp\left(-\frac{\lambda+x}{2}\right)
\sum_{j=0}^{\infty}\frac{(\sqrt{\lambda x})^{2j}}{2j!}
\frac{2j!}{2^{2j}j!(k/2+j-1)\times \cdots \times(k/2)} \cr
&\leq & \frac{x^{k/2-1}}{\Gamma(k/2)2^{k/2}}\exp\left(-\frac{\lambda+x}{2}\right)
\sum_{j=0}^{\infty}\frac{( \sqrt{\lambda x})^{2j} }{2j! } \cr
&\leq& \frac{x^{k/2-1}}{\Gamma(k/2)2^{k/2}}\exp\left\{-\frac{(\sqrt{\lambda}-\sqrt{x})^2}{2}\right\}
\end{eqnarray}

Let $\Lambda_i(x):=\int_0^\infty \lambda^if_{k,\lambda}(x)dG(\lambda)$. By (\ref{bndnc}),
\[
\Lambda_i(x)\leq \frac{x^{k/2-1}}{\Gamma(k/2)2^{k/2}}\int_0^\infty\lambda^i\exp\left\{-\frac{(\sqrt{\lambda}-\sqrt{x})^2}{2}\right\}dG(\lambda).
\]
The integrand of the integral above
\[\lim_{\lambda\rightarrow\infty}\lambda^i\exp\left\{-\frac{(\sqrt{\lambda}-\sqrt{x})^2}{2}\right\}=0\]
because $\lambda^i$ grows at a polynomial rate and $ \exp\{-(\sqrt{\lambda}-\sqrt{x})^2/2\}\downarrow 0$ at an exponential rate as $\lambda\rightarrow\infty$. Since the integrand is continuous, it must be bounded over $[0,\infty)$. This implies that $0<\Lambda_i(x)<\infty$ for each $x>0$. The proof is completed by observing
\[ E_k(\lambda|x)=\frac{\Lambda_1(x)}{\Lambda_0(x)}, {\rm and}\;\; {\rm var}_k(\lambda|x)=\frac{\Lambda_2(x)}{\Lambda_0(x)}-\left(\frac{\Lambda_1(x)}{\Lambda_0(x)}\right)^2.\]

\noindent{\Large \bf Acknowledgement}\\
We thank the Editor, Associate Editor and reviewers for helpful comments and suggestions. The initial ideas of this work were conceived while Inchi Hu was on sabbatical leave in Taiwan 2016/2017. The hospitality of Professors Mei-Hui Guo and Mong-Na Lo Huang of National Sun Yat Sen University, Professor Cheng-Der Fuh of National Central University, Professors Yi-Ching Yao and Feng-Shun Chai of Institute of Statistical Science, Academia Sinica, and Professor Chii-Ruey Hwang of Institute of Mathematics, Academia Sinica is gratefully acknowledged. The research of Inchi Hu is supported by Grant SGI19BM01 and Lilun Du by Hong Kong RGC ECS26301216.


\end{document}